\documentclass{article}
\usepackage{spconf,amsmath,graphicx,hyperref, bm}
\usepackage{xcolor}
\usepackage{amssymb,amsfonts, mathtools}

\usepackage{fancyhdr}

\fancypagestyle{firstpage}{%
  \fancyhf{}

  \fancyfoot[C]{%
    \fbox{%
      \begin{minipage}{2\columnwidth}
        \footnotesize
        This paper has been accepted to the IEEE International Conference on Acoustics, Speech, and Signal Processing (ICASSP) 2026.\\
        © 2026 IEEE. Personal use of this material is permitted. Permission from IEEE must be obtained for all other uses, in any current or future media, including reprinting/republishing this material for advertising or promotional purposes, creating new collective works, for resale or redistribution to servers or lists, or reuse of any copyrighted component of this work in other works.
      \end{minipage}
    }%
  }
}

\title{Exterior sound field estimation based on physics-constrained kernel}
\name{Juliano G. C. Ribeiro, Ryo Matsuda, Jorge Trevino}
\address{Yamaha Corporation, Hamamatsu, Japan}
\begin{document}
\ninept
\maketitle
\thispagestyle{firstpage}
\begin{abstract}
  Exterior sound field interpolation is a challenging problem that often requires specific array configurations and prior knowledge on the source conditions. We propose an interpolation method based on Gaussian processes using a point source reproducing kernel with a trainable inner product formulation made to fit exterior sound fields. While this estimation does not have a closed formula, it allows for the definition of a flexible estimator that is not restricted by microphone distribution and attenuates higher harmonic orders automatically with parameters directly optimized from the recordings, meaning an arbitrary distribution of microphones can be used. The proposed kernel estimator is compared in simulated experiments to the conventional method using spherical wave functions and an established physics-informed machine learning model, achieving lower interpolation error by approximately $2~\mathrm{dB}$ on average within the analyzed frequencies of $100~\mathrm{Hz}$ and $2.5~\mathrm{kHz}$ and reconstructing the ground truth sound field more consistently within the target region.
\end{abstract}
\section{Introduction}
\label{sec:intro}

The interpolation of a sound field based on microphone recordings is a foundational problem in acoustic signal processing~\cite{Williams:FourierAcoust,Rafaely:FundSphArrayProc}. The interpolation of exterior sound fields, \textit{i.e.} sound fields in regions surrounding a source boundary, have several practical applications such as wave field synthesis~\cite{Okamoto:WASPAA2021}, active noise control~\cite{Arikawa:ICA2022}, and sound field reproduction~\cite{Fazi:AmbiSymp2009,Ueno:IEEE_ACM_J_ASLP2019}. There are also related problems to exterior sound field interpolation that benefit from a more thorough understanding of the subject, such as sound field recordings in the presence of scatterers~\cite{Koyama:WASPAA2023,Matsuda:WASPAA2025}, and sound field separation~\cite{PEZZOLI:AA2024, Bi:JASA2008}. However, the analysis and approximation of exterior sound field functions is challenging and often requires the use of additional information~\cite{Nguyen:IEEE_J_IM2025}, or requires the exploration of locality~\cite{Ren:WASPAA2021}.

Utilizing neural networks, it is possible to approximate interior sound fields with entirely data-driven approaches~\cite{Pezzoli:Sensors2022} or physics-informed neural networks~\cite{Raissi:CompPhys2019} in order to include acoustic properties into training~\cite{Karakonstantis:JASA2024, Chen:APSIPA2023}. However, the sound pressure field in point-source models diverges at the source location~\cite{Williams:FourierAcoust, Rudin:RealAndComplexAnalysis}, which can make neural networks difficult to train unless they incorporate properties of the exterior sound field in their design~\cite{Bi:JASMP2024}. Another approach is to approximate the sound field as a combination of wave function solutions of the Helmholtz equation~\cite{Poletti:J_AES_2005, Ueno:FTSP2025}. However, these systems tend to be rather sensitive to regularization and microphone distribution. The spherical wave function expansion for the interior sound field can be extended to a closed form expression without truncation~\cite{Ueno:IEEE_SPL2018} using the kernel method, with the option to improve accuracy using learnable kernel functions~\cite{Ribeiro:ICASSP2022}.

In this work, we propose a Gaussian process regression model with a multipole point source kernel that employs a weighted inner product. This inner product uses a parametric weight, meaning the attenuation of higher order components is automatically decided by two parameters. The kernel is also based on the exterior wave function solutions of the inhomogeneous Helmholtz equation, meaning they satisfy the physical constraint by default and their optimization can be entirely data-driven. Gaussian processes have also been applied in problems involving source information~\cite{Matsuda:JASA2025}, making Gaussian process regression a natural choice. We compare this proposed methodology with the more standard spherical wave function expansion as well as with an established learnable model~\cite{Bi:JASMP2024} in numerical simulations assuming both a spherical array distribution and a completely random distribution. 

\section{Problem statement}
\label{sec:probstat}

\begin{figure}[t]
\centering
\begin{minipage}[b]{0.7\linewidth}
  \centering
  \centerline{\includegraphics[width=5cm]{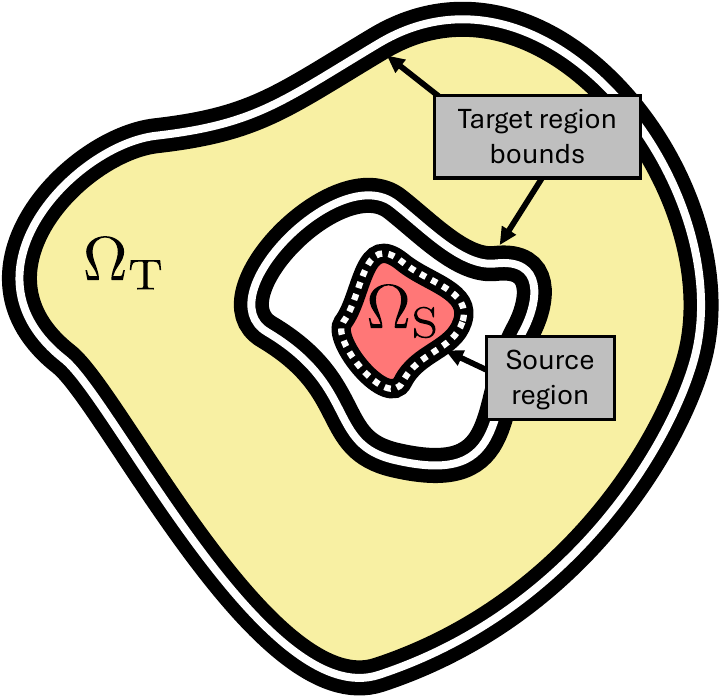}}
\end{minipage}
\vspace{-0.3cm}
\caption{Example of the problem statement showing the bounds of the target region and the source region enclosed by it.}
\label{fig:ProbStat}
\end{figure}

Suppose a space $\Omega\subset \mathbb{R}^3$ enclosing a source region $\Omega_\mathrm{S} \subset \Omega$ and a target region $\Omega_\mathrm{T} \subset \Omega$. Suppose also that $\Omega_\mathrm{T}$ has interior and exterior boundaries such that $\Omega_\mathrm{S}$ is simply connected and is contained by the interior boundary of $\Omega_\mathrm{T}$. This scheme can be seen in Fig.~\ref{fig:ProbStat}.

The frequency domain sound field $u:\Omega \times \mathbb{R} \rightarrow \mathbb{C}$ is an exterior sound field if it respects the Helmholtz equation~\cite{Williams:FourierAcoust} outside $\Omega_\mathrm{S}$ and satisfies the Sommerfeld radiation conditions~\cite{Williams:FourierAcoust, Ueno:FTSP2025} for distances remote from $\Omega_\mathrm{S}$, meaning
\begin{align}
  &\nabla^2 u(\mathbf{r}, k) + k^2 u(\mathbf{r}, k) = 0, \forall \mathbf{r}\in \Omega / \Omega_\mathrm{S},\\
  &\underset{\|\mathbf{r}\| \rightarrow \infty}{\mathrm{lim}} \|\mathbf{r}\| \left ( \frac{\partial u}{\partial \|\mathbf{r}\|} - \mathrm{i} k \right )u(\mathbf{r}) = 0,
\end{align}
where $\mathbf{r}$ is the position and $k$ is the wave number. Given the discussed methods in this work operate in the frequency domain and each frequency bin is considered to be independent, $k$ will be ommitted from arguments for simplicity unless explicitly needed. Our objective is to approximate $u$ based on recordings taken from $M$ microphones located in positions $\{\mathbf{r}_m\}_{m=1}^{M}\subset \Omega_\mathrm{T}$.

\section{Conventional methods}
\label{sec:convmet}

\subsection{Spherical wave function expansion}

An exterior sound field function can be represented by a superposition of solutions to the inhomogeneous Helmholtz equation~\cite{Williams:FourierAcoust}, here given as
\begin{equation}
  \psi_{\nu, \mu}(\mathbf{r}) = h_\nu(k\|\mathbf{r}\|) Y_{\nu}^\mu\left ( \frac{\mathbf{r}}{\|\mathbf{r}\|}\right ),
\end{equation}
where $h_\nu$ is the spherical Hankel function of order $\nu \in \mathbb{N}$ of the first kind, relating to an outgoing sound field~\cite{Ueno:FTSP2025}, and $Y_\nu^\mu$ is the spherical harmonic function of order $\nu$ and mode $\mu\in \{-\nu, -\nu+1, \dots, \nu-1, \nu\}$. Then, from \cite{Poletti:J_AES_2005}, the exterior sound field can be approximated as
\begin{equation}\label{eq:SWF_estimator}
  \hat{u}_\mathrm{SWF}(\mathbf{r}) = \sum_{\nu = 0}^{\nu_\mathrm{SWF}} \sum_{\mu = -\nu}^{\nu} \mathring{u}_{\nu, \mu} \psi_{\nu, \mu}(\mathbf{r}),
\end{equation}
where $\nu_\mathrm{SWF}$ is the truncation order and $\mathring{\mathbf{u}}\in \mathbb{C}^{(\nu_\mathrm{SWF}+1)^2}$ is the vector of coefficients $\mathring{u}_{\nu, \mu}$. The truncation order is assumed to be the maximum value of $\nu_\mathrm{SWF}$ such that $(\nu_\mathrm{SWF}+1)^2 \leq M$. We can learn the coefficents by solving the following optimization problem:
\begin{align}
  \underset{\mathring{\mathbf{u}} \in \mathbb{C}^{(\nu_\mathrm{SWF} + 1)^2}}{\mathrm{minimize}} & (\mathbf{s} - \mathbf{\Psi} \mathring{\mathbf{u}})^\mathsf{H} \mathbf{W} (\mathbf{s} - \mathbf{\Psi} \mathring{\mathbf{u}}) + \lambda_\mathrm{SWF} \mathring{\mathbf{u}}^\mathsf{H} \mathbf{D} \mathring{\mathbf{u}},
\end{align}
where $\mathbf{W}$ is a diagonal matrix where the entries are the quadrature weights corresponding to each position $\mathbf{r}_m$ in spherical coordinates, $\mathbf{D}$ is a diagonal matrix where $[\mathbf{D}]_{\nu^2 + \nu + \mu + 1, \nu^2 + \nu + \mu + 1}= \nu^2 + \nu + 1$~\cite{Duraiswami:ICASSP2004}, $\lambda_\mathrm{SWF}\in \mathbb{R}_+$ is the regularization constant, $\mathbf{s}\in \mathbb{C}^M$ is the matrix of recordings for each microphone, and $\mathbf{\Psi} \in \mathbb{C}^{M \times (\nu_\mathrm{SWF}+1)^2}$ is the matrix of spherical wave function evaluations $[\mathbf{\Psi}]_{m, \nu^2 + \nu + \mu +1} = \psi_{\nu, \mu}(\mathbf{r}_m)$. The optimal value for the coefficients is given as
\begin{equation}
  \mathring{\mathbf{u}}_\mathrm{opt} = \left ( \mathbf{\Psi}^\mathsf{H} \mathbf{W} \mathbf{\Psi} + \lambda_\mathrm{SWF} \mathbf{D} \right )^{-1}(\mathbf{\Psi}^\mathsf{H} \mathbf{W} \mathbf{s}).\label{eq:SWF_opt}
\end{equation}

\subsection{Point neuron network}
In~\cite{Bi:JASMP2024}, the authors describe an estimator that is physics-based and uses fully learnable centers and weights, the point neuron network. The estimator $\hat{u}_\mathrm{PNN}$ is
\begin{equation}
  \hat{u}_\mathrm{PNN}(\mathbf{r}) = \sum_{n = 1}^{N_\mathrm{PNN}} \eta_n \|\mathbf{v}_n\| \frac{\mathrm{e}^{\mathrm{i}k (\|\mathbf{r} - \mathbf{v}_n\| - \|\mathbf{v}_n\|)}}{4\pi \|\mathbf{r} - \mathbf{v}_n\|},
\end{equation}
where $N_\mathrm{PNN}$ is the number of sources, $\bm{\eta} = [\eta_n]_{n=1}^{N_\mathrm{PNN}} \subset \mathbb{C}^{N_\mathrm{PNN}}$ are the weights associated with each point neuron and $\mathbf{v}_n \in \mathbb{R}^3$ are the positions of the point neurons. Then, we can obtain the optimal parameters to approximate $u$ by optimizing
\begin{align}
  \underset{\bm{\eta} \in \mathbb{C}^{N_\mathrm{PNN}}, \mathbf{v}_n \subset \mathbb{B}\  \forall n}{\mathrm{minimize}} \sum_{m=1}^{M} |\hat{u}_\mathrm{PNN}(\mathbf{r}_m) - s_m|^2 + \lambda_\mathrm{PNN} \|\bm{\eta} \|_1,
\end{align}
where $\mathbb{B}$ is the interior bound of $\Omega_\mathrm{T}$, $\|\cdot \|_1$ is the $L_1$ norm, used here to induce sparsity, and $\lambda_\mathrm{PNN}\in \mathbb{R}_+$ is a regularization constant.

\section{Proposed method}
\label{sec:proposed}

\begin{figure}[t]
\centering
\begin{minipage}[b]{0.7\linewidth}
  \centering
  \centerline{\includegraphics[width=7.5cm]{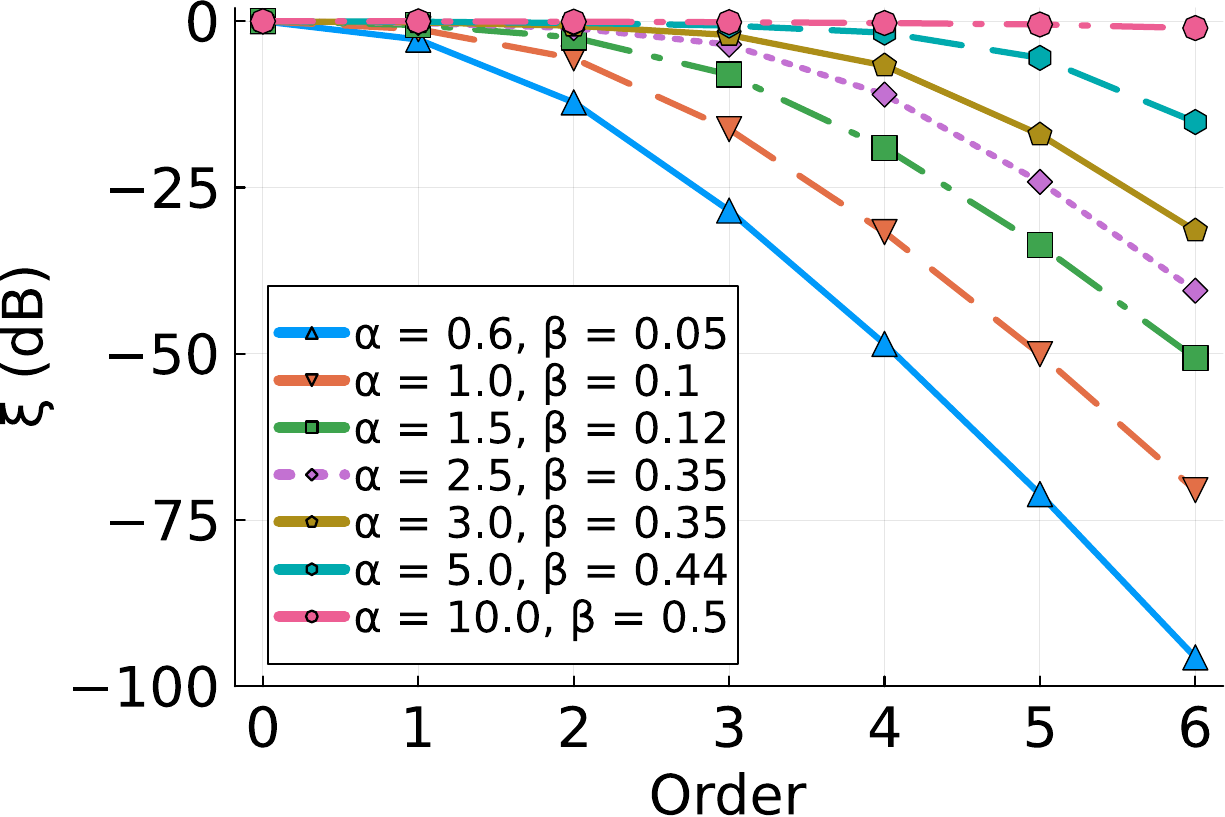}}
\end{minipage}
\hspace{0.5cm}
\vspace{-0.3cm}
\caption{Proposed method mode attenuation scheme as function of weight parameters $\alpha$ and $\beta$.}
\label{fig:Attenuation}
\vspace{-0.3cm}
\end{figure}
The proposed method aims to satisfy the physical constraints by using the exterior sound field spherical wave functions to define a reproducing kernel Hilbert space (RKHS)~\cite{Rudin:FunctionalAnalysis} similarly to \cite{Ueno:IEEE_SPL2018}. Therefore, we define our RKHS with these wave functions as a central feature and divergences are addressed using a parametric attenuation function.

\subsection{RKHS definition}

\begin{figure*}[ht]
\hspace{0.33cm}
\begin{minipage}[b]{.45\linewidth}
  \centering
  \centerline{\includegraphics[width=8.8cm]{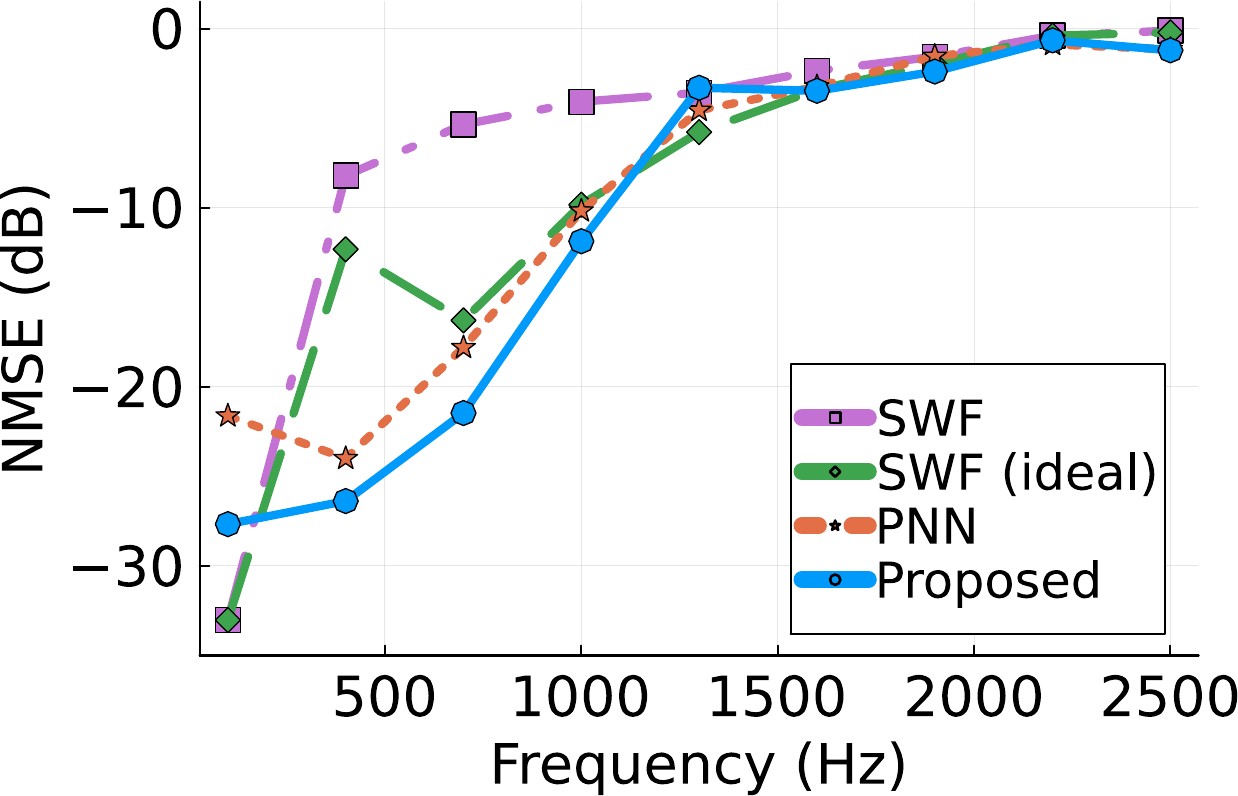}}
  \centerline{\hspace{0.5cm}(a) t-design array}\medskip
\end{minipage}
\hspace{0.79cm}
\begin{minipage}[b]{.45\linewidth}
  \centering
  \centerline{\includegraphics[width=8.8cm]{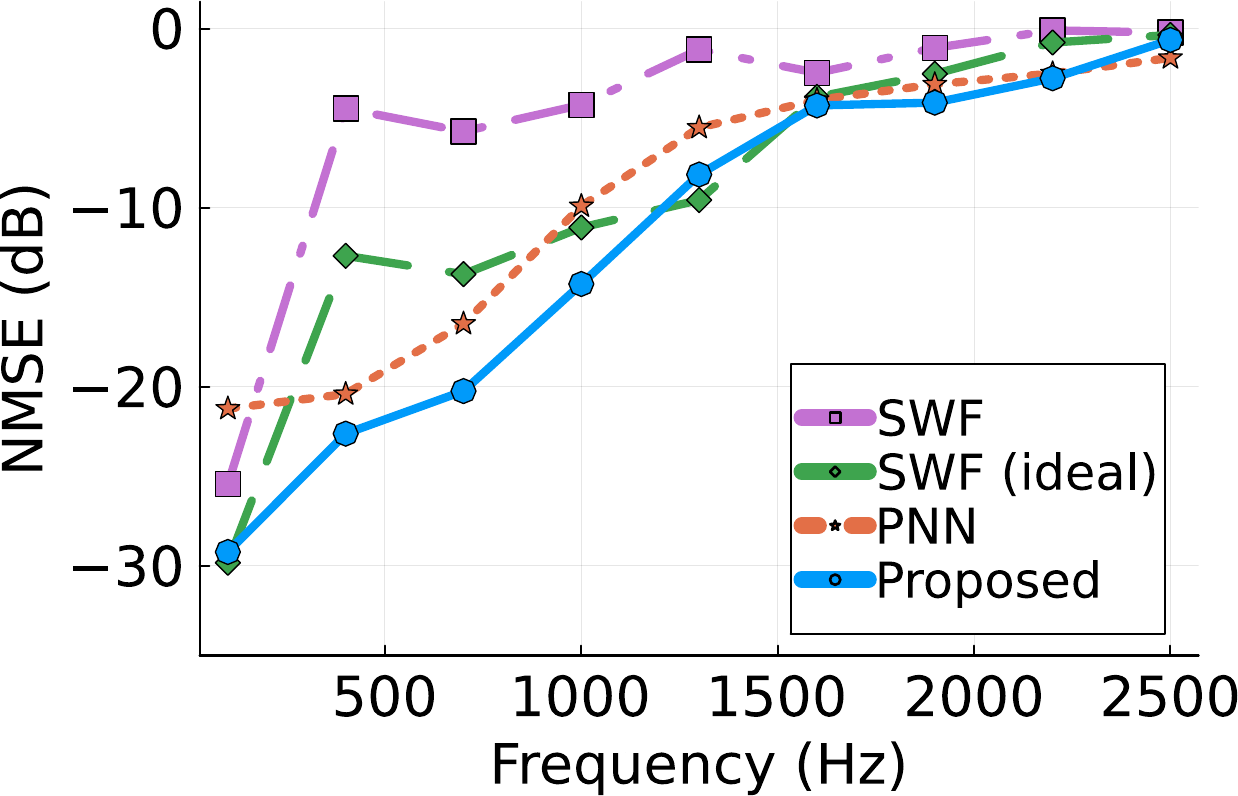}}
  \centerline{\hspace{0.5cm}(b) random array}\medskip
\end{minipage}
\vspace{-0.4cm}
\caption{NMSE as a function of frequency for each method.}
\label{fig:NMSE}
\vspace{-0.3cm}
\end{figure*}

The first step to properly define the RKHS is to create an inner-product that will be convergent despite the presence of poles. We opted to develop an inner product using a weighted integral formulation
\begin{equation}
  \langle u_1, u_2 \rangle_{w} = \int_{\mathbb{R}^3} w(\mathbf{r}) \overline{u_1(\mathbf{r})} u_2(\mathbf{r}) \mathrm{d}\mathbf{r}
\end{equation}
where $w:\mathbb{R}^3 \rightarrow \mathbb{R}_+$ is the weight function and $\overline{(\cdot)}$ is the complex conjugate. Given the spherical harmonic functions have no poles, we can deduce the issues relate to the fact the spherical Hankel function of order $\nu \in \mathbb{N}$, $h_\nu$, has a pole of order $\nu+1$. For that reason, we assume the weight function to be a radial function $w(\mathbf{r}) = w_0(\|\mathbf{r}\|)$ such that $w_0:\mathbb{R}_+ \rightarrow \mathbb{R}_+$ and
\begin{equation}
  \int_{0}^{\infty} \frac{w_0(r)}{r^{2(\nu+1)}} < \infty,\ \forall \nu \in \mathbb{N}.\label{eq:conv_cond}
\end{equation}

Since no finite order of zeros will satisfy convergence for all orders, the weight has to be carefully defined, such as the family of weight functions:
\begin{equation}
  w(\mathbf{r}) = k\mathrm{e}^{-\left (\frac{\alpha}{k\|\mathbf{r}\|}\right )^{1/\beta}},
\end{equation}
for which the condition~\eqref{eq:conv_cond} is satisfied as long as $\alpha > 0$ and $\beta > 0$. By performing a simple change of variable we can show that
\begin{align}
  \|\psi_{\nu, \mu}\|^2 &= \langle \psi_{\nu, \mu}(\mathbf{r}), \psi_{\nu, \mu}(\mathbf{r})\rangle\notag\ \mathrm{and} \\
  \langle \psi_{\nu, \mu}(\mathbf{r}), \psi_{\nu^\prime, \mu\prime}(\mathbf{r})\rangle &= \begin{cases}0,\ \nu \neq \nu^\prime\ \mathrm{or}\ \mu \neq \mu^\prime\\ \int_{\mathbb{R}^3} w_0(\|\mathbf{r}\|) |\psi_{\nu, \mu}(\mathbf{r})|^2 \mathrm{d}\mathbf{r}\end{cases},
\end{align}
meaning $\|\psi_{\nu, \mu}\|<\infty$, properly defining our inner product as
\begin{equation}
\langle u_1, u_2 \rangle_{\alpha, \beta} = \int_{\mathbb{R}^3}  k\mathrm{e}^{-\left (\frac{\alpha}{k\|\mathbf{r}\|}\right )^{1/\beta}} \overline{u_1(\mathbf{r})} u_2(\mathbf{r}) \mathrm{d}\mathbf{r}.
\end{equation}

Furthermore, as the weight is only a function of the distance, it is isotropic and the norms of wave functions of the same order are equal:
\begin{align}
  \|\psi_{\nu, \mu}\|^2 &= \langle \psi_{\nu, \mu}, \psi_{\nu, \mu}\rangle\notag \\
  &=\int_{\mathbb{R}^3} k\mathrm{e}^{\left (\frac{\alpha}{k\|\mathbf{r}\|}\right )^{1/\beta}}|h_\nu(k\|\mathbf{r}\|)|^2 \left | Y_\nu^\mu\left ( \frac{\mathbf{r}}{\|\mathbf{r}\|} \right ) \right |^2 \mathrm{d}\mathbf{r}\notag \\
  &=\int_0^\infty \mathrm{e}^{-\left (\frac{\alpha}{r}\right )^{1/\beta}}|h_\nu(r)|^2\mathrm{d}r,
\end{align}
which allows us to define the attenuation of each order that enables the definition of a RKHS as
\begin{equation}
  \xi_\nu(\alpha, \beta) = \frac{1}{\int_0^\infty \mathrm{e}^{-\left (\frac{\alpha}{r}\right )^{1/\beta}}|h_\nu(r)|^2\mathrm{d}r},
\end{equation}
which can be seen in Fig.~\ref{fig:Attenuation}. By optimizing $\alpha$ and $\beta$, it is possible to automatically tune the mode cut off based on the observed data.

Finally, all that is left is to create the RKHS proper. We begin by defining the Hilbert space below:
\begin{equation}
  \mathcal{H} =\left \{u = \sum_{\nu = 0}^{\nu_\mathrm{KRR}} \sum_{\mu = -\nu}^{\mu} \tilde{u}_{\nu, \mu} \psi_{\nu, \mu} : \sum_{\nu \in \mathbb{N}} \frac{|\tilde{u}_{\nu, \mu}|^2}{\xi_\nu(\alpha, \beta)} <\infty \right \},
\end{equation}
where $\nu_\mathrm{KRR}$ is the truncation order. This space inherits its completeness from the completeness of $\ell_2$ spaces~\cite{Rudin:FunctionalAnalysis} and the following kernel function
\begin{equation}
  \kappa(\mathbf{r}, \mathbf{r}^\prime; \alpha, \beta) = \sum_{\nu=0}^{\nu_\mathrm{KRR}}\sum_{\mu = -\nu}^{\nu} \xi_\nu(\alpha, \beta) \psi_{\nu, \mu}(\mathbf{r})\overline{\psi_{\nu, \mu}(\mathbf{r}^\prime)}
\end{equation}
is a reproducing kernel for any $\nu_\mathrm{KRR} \in \mathbb{N}\cup \{+\infty\}$. This kernel does not have a known closed form for $\nu_\mathrm{KRR} = \infty$. However, unlike approaches that aim to approximate data using wave functions directly, the order of truncation is not tied to the number of microphones. In our case, the truncation order was fixed at $\nu_\mathrm{KRR} = 20$.

\begin{figure*}[t]
\hspace{0.5cm}
\begin{minipage}[b]{.22\linewidth}
  \centering
  \centerline{\includegraphics[width=3.5cm]{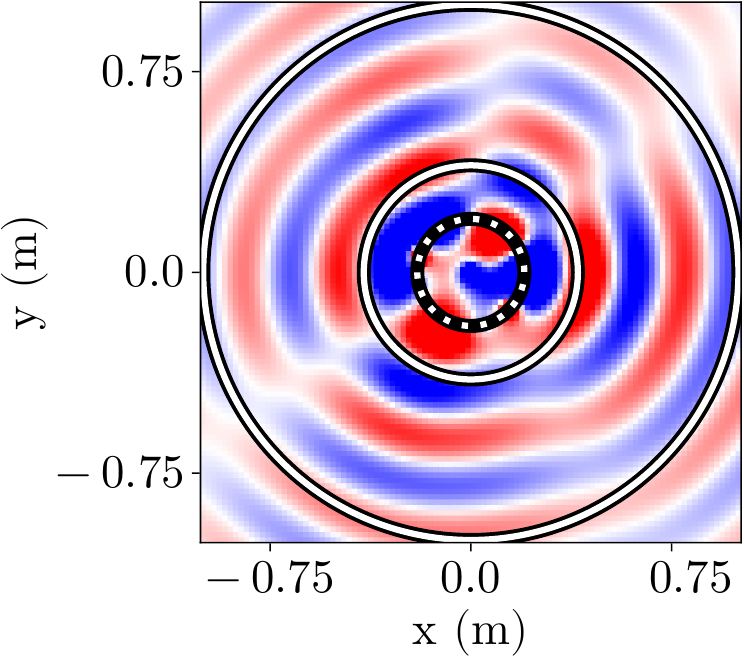}}
  \centerline{\hspace{0.7cm}(a) Ground truth}\medskip
\end{minipage}
\begin{minipage}[b]{.22\linewidth}
  \centering
  \centerline{\includegraphics[width=3.5cm]{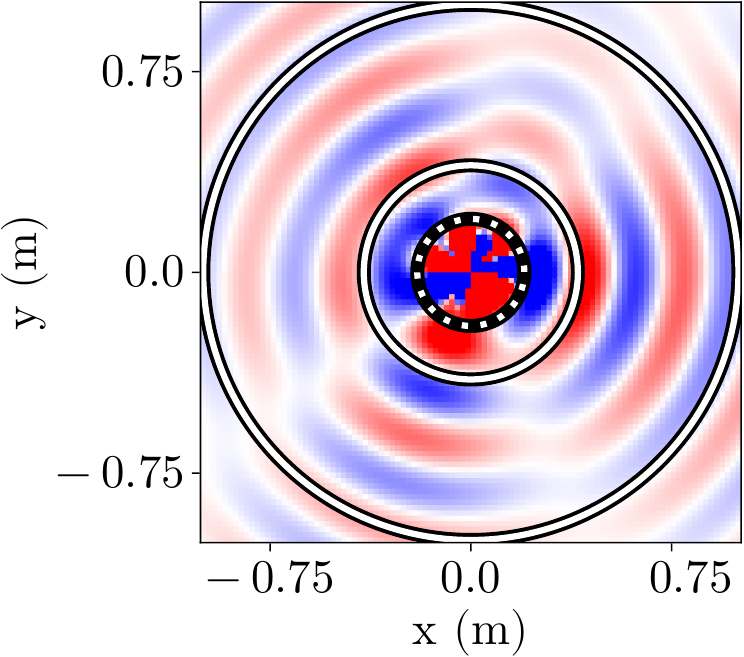}}
  \centerline{\hspace{0.7cm}(b) \textbf{SWF}}\medskip
\end{minipage}
\begin{minipage}[b]{0.22\linewidth}
  \centering
  \centerline{\includegraphics[width=3.5cm]{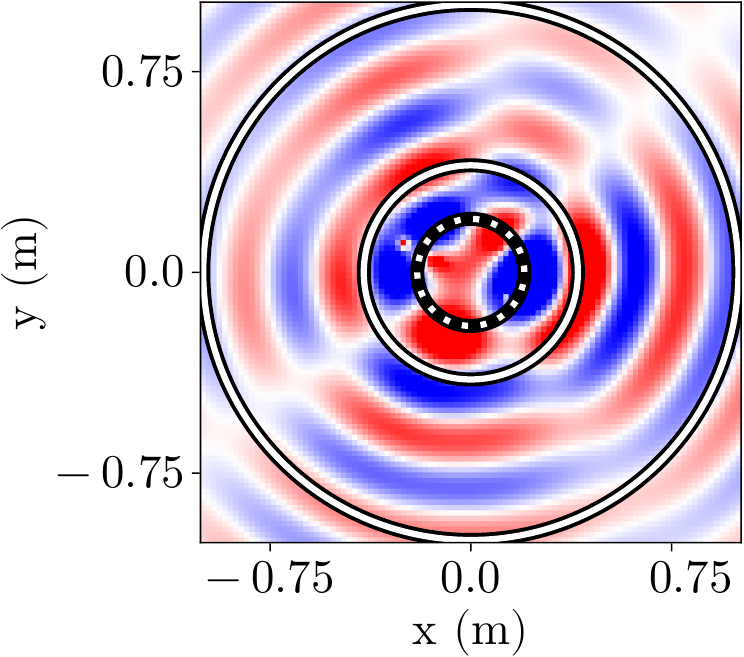}}
  \centerline{\hspace{0.7cm}(c) \textbf{PNN}}\medskip
\end{minipage}
\begin{minipage}[b]{0.22\linewidth}
  \centering
  \centerline{\includegraphics[width=3.5cm]{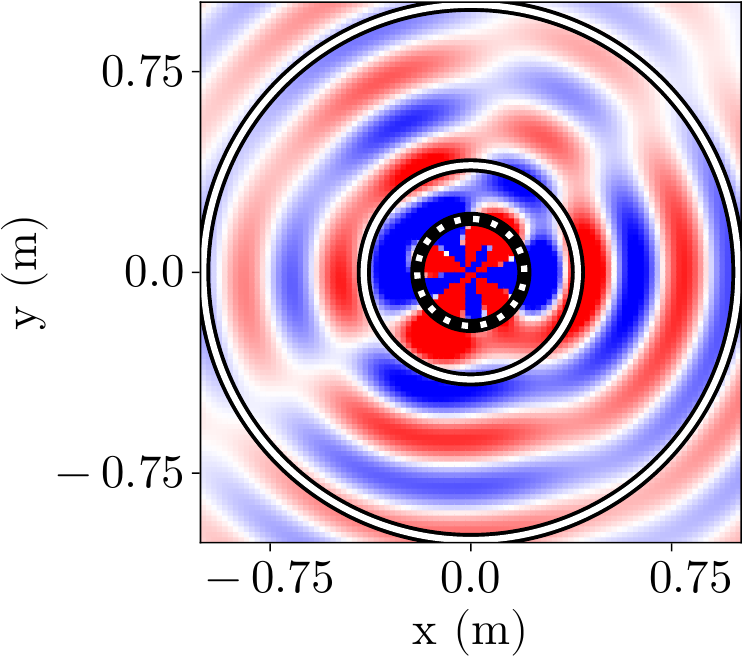}}
  \centerline{\hspace{0.7cm}(d) \textbf{Proposed}}\medskip
\end{minipage}
\hspace{-0.6cm}
\begin{minipage}[b]{0.1\linewidth}
  \centering
  \centerline{\includegraphics[height=2.7cm, width = 1.cm]{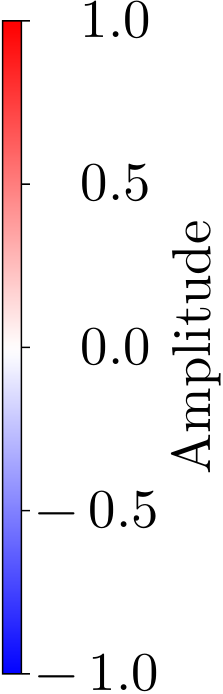}}
 \vspace{1.1cm}
\end{minipage}
\vspace{-0.9cm}
\caption{Real part of the ground truth as well as estimated sound fields. The white circles represent the bounds of the target region, while the dotted circle represents the bounds of the source region.}
\label{fig:VIS}
\vspace{0.2cm}
\hspace{4.495cm}
\begin{minipage}[b]{.22\linewidth}
  \centering
  \centerline{\includegraphics[width=3.5cm]{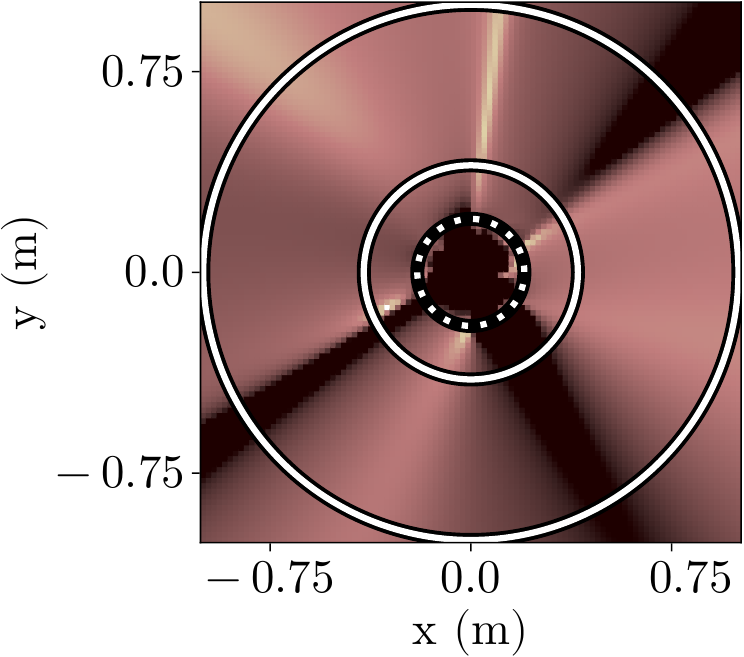}}
  \centerline{\hspace{0.7cm}(b) \textbf{SWF}}\medskip
\end{minipage}
\begin{minipage}[b]{0.22\linewidth}
  \centering
  \centerline{\includegraphics[width=3.5cm]{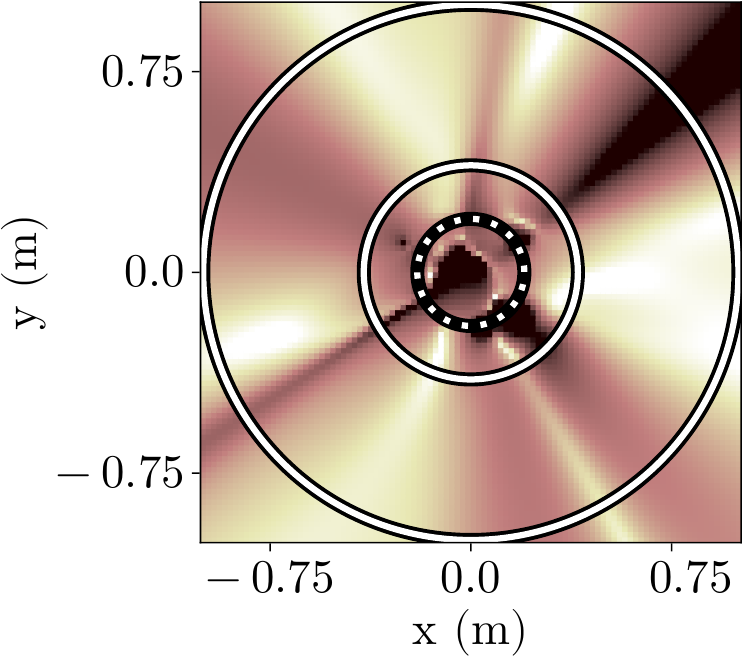}}
  \centerline{\hspace{0.7cm}(c) \textbf{PNN}}\medskip
\end{minipage}
\begin{minipage}[b]{0.22\linewidth}
  \centering
  \centerline{\includegraphics[width=3.5cm]{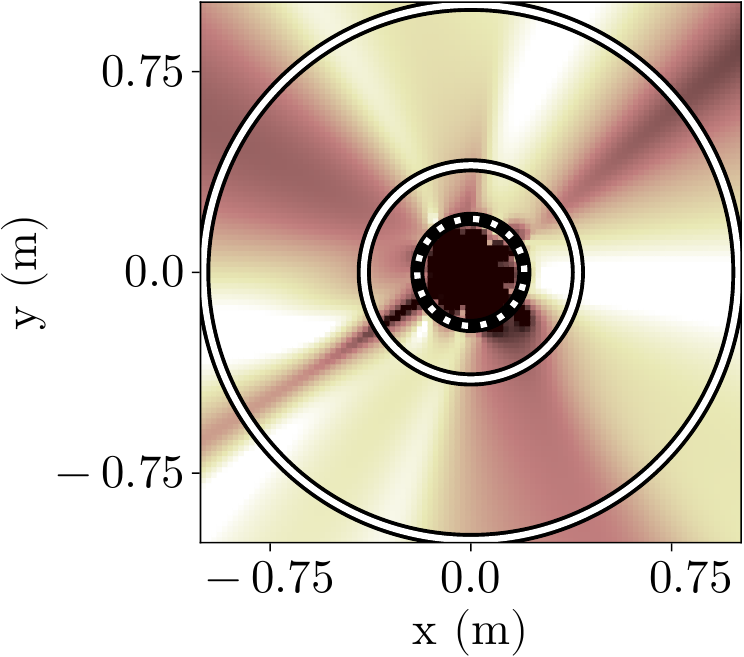}}
  \centerline{\hspace{0.7cm}(d) \textbf{Proposed}}\medskip
\end{minipage}
\hspace{-0.6cm}
\begin{minipage}[b]{0.1\linewidth}
  \centering
  \centerline{\includegraphics[height=2.7cm, width = 1.cm]{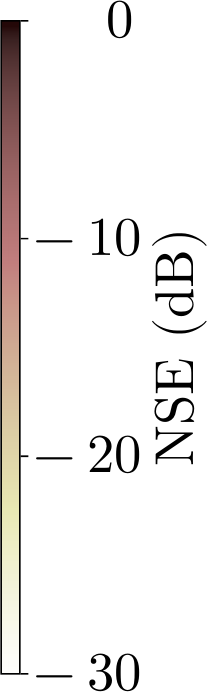}}
 \vspace{1.1cm}
\end{minipage}
\hspace{0.6cm}
\vspace{-0.9cm}
\caption{NSE distribution of the estimated sound field reconstruction attempts for each method.}
\label{fig:NSE}
\vspace{-0.3cm}
\end{figure*}

\subsection{Model derivation and optimization}

With the kernel model defined, the estimator $\hat{u}_\mathrm{KRR}$ takes the form
\begin{equation}
  \hat{u}_{\mathrm{KRR}}(\mathbf{r}) = \sum_{m=1}^M a_m \kappa_{\alpha, \beta}(\mathbf{r}, \mathbf{r}_m).
\end{equation}
where the coefficient vector $\mathbf{a} = [a_m]_{m=1}^M \in \mathbb{C}^M$ can be estimated using kernel ridge regression (KRR)~\cite{Murphy:ML} to be
\begin{align}
  \hat{\mathbf{a}} &= \underset{\mathbf{a} \in \mathbb{C}^M}{\mathrm{argmin}} \| \mathbf{s} - \mathbf{K} \mathbf{a}\|^2 + \lambda_\mathrm{KRR} \mathbf{a}^\mathsf{H} \mathbf{K} \mathbf{a}\notag \\
  &= \left ( \mathbf{K} + \lambda_\mathrm{KRR} \mathbf{I} \right )^{-1} \mathbf{s},
\end{align}
where $\mathbf{K} = [\kappa_{\alpha, \beta}(\mathbf{r}_{m_1}, \mathbf{r}_{m_2})]_{m_1 \in \{1, 2, \dots, M\}, m_2 \in \{1, 2, \dots, M\}} \in \mathbb{C}^{M \times M}$ is the Gram matrix and $\lambda_\mathrm{KRR} \in \mathbb{R}_+$ is the regularization constant.

The model is optimized with constraints using Gaussian process regression (GPR) and the logarithmic likelihood assuming Gaussian priors~\cite{Murphy:ML}, with a stability criterion using the condition number of the Gram matrix:
\begin{align}
  \hat{\alpha}, \hat{\beta} = &\underset{\alpha, \beta}{\mathrm{argmin}}\ \splitfrac{\mathbf{s}^\mathsf{H}(\mathbf{K} + \lambda_\mathrm{KRR} \mathbf{I})^{-1} \mathbf{s} + \mathrm{logdet}(\mathbf{K} + \lambda_\mathrm{KRR}\mathbf{I})}{+\lambda_\mathrm{cond} \mathrm{log}(\mathrm{cond}(\mathbf{K}+ \lambda_\mathrm{KRR} \mathbf{I}))\hspace{2cm}}\notag \\
  &\mathrm{s.\ t.\ }\Delta_\mathrm{min} \leq \alpha - \beta \leq \Delta_\mathrm{max}\ \mathrm{and}\ \beta \in [\beta_\mathrm{min}, \beta_\mathrm{max}],\label{eq:opt}
\end{align}
where $\lambda_\mathrm{cond} \in \mathbb{R}_+$ is the regularization constant for the condition number, $\Delta_\mathrm{min}>0$ and $\Delta_\mathrm{max}>\Delta_\mathrm{min}$ are the minimum and maximum separation between the parameters while $\beta_\mathrm{min}$ and $\beta_\mathrm{max}$ are the minimum and maximum values for $\beta$, respectively.

\section{Numerical experiments}

The conventional methods were compared to the proposed method in numerical simulations using several monopole sources. The source region $\Omega_\mathrm{S}$ was considered to be a sphere of radius $R_\mathrm{s} = 0.2~\mathrm{m}$, while $\Omega_\mathrm{T}$ was a spherical shell with internal radius $R_\mathrm{in} = 0.4~\mathrm{m}$ and external radius $R_\mathrm{out} = 1.0~\mathrm{m}$, both regions centered at the origin. We distributed $27$ monopole sources in $\Omega_\mathrm{S}$, with $26$ on the surface distributed according to the spherical t-design grid of order 6~\cite{Sloane:T_design}, as well as $1$ center source, and each source was assigned a random coefficient following a complex Gaussian distribution $\mathcal{N}_\mathbb{C}(0, 1)$.

\subsection{Evaluation metrics}

We evaluated the models in two distributions for microphone positions: a spherical t-design of order $9$ with a radius of $0.81~\mathrm{m}$, resulting in 48 points, and another with 50 uniformly sampled points within $\Omega_\mathrm{T}$. To the samples of each array, random Gaussian noise was added so the signal-to-noise ratio would be $20~\mathrm{dB}$.

The first criterion for evaluation was a selection of $M^\prime = 500$ random test points $\{\mathbf{r}^\prime_{m^\prime}\}_{m^\prime=1}^{M^\prime}$ distinct from the microphone positions and sampled uniformly from $\Omega_\mathrm{T}$, meant to measure the overall accuracy of each method in the target region using the normalized mean square error ($\mathrm{NMSE}$):
\begin{equation}
  \mathrm{NMSE}(u, \hat{u}) = 10\log_{10}\left (\frac{\sum_{m^\prime = 1}^{M^\prime} |u(\mathbf{r}^\prime_{m^\prime}) - \hat{u}(\mathbf{r}^\prime_{m^\prime})|^2}{\sum_{m^\prime = 1}^{M^\prime} |u(\mathbf{r}^\prime_{m^\prime})|^2} \right ).
\end{equation}

For the second criterion, $10000$ points were distributed in the plane $z=0$ around the source region in a square of side $2~\mathrm{m}$. For each point in that plane, each method tried to reconstruct the sound field in that position for a frequency of $1~\mathrm{kHz}$. For a more qualitative test, we also calculated the normalized square error ($\mathrm{NSE}$):
\begin{equation}
  \mathrm{NSE}(\mathbf{r}, u, \hat{u}) = 20\log_{10} \left ( \frac{|u(\mathbf{r}) - \hat{u}(\mathbf{r})|}{|u(\mathbf{r})|} \right ).
\end{equation}

\subsection{Training of the models}
For the proposed method, henceforth \textbf{Proposed}, optimization in \eqref{eq:opt} is performed initially with a random value for regularization constant~$\lambda_\mathrm{KRR}$ chosen uniformly in logarithmic scale as $\log_{10}\lambda_\mathrm{KRR} \in [-3, 1]$ as well as $\lambda_\mathrm{cond} = 0.0075$, chosen empirically. The optimization was performed using established routines~\cite{Dixit:Zenodo2023, Zygote.jl-2018} and the constraints were assigned as $\Delta_\mathrm{min} = 1$, $\Delta_\mathrm{max} = 100$, $\beta_\mathrm{min} = 10^{-4}$, and $\beta_\mathrm{max} = 5$. After optimization was performed, we executed a grid search for a more adequate value of $\lambda_\mathrm{KRR}$, also in logarithmic scale, such that $-10 \leq \log_{10}\lambda_\mathrm{KRR} \leq 2$ with a $0.25$ step size. The loss criterion for this case was leave-one-out cross validation (LOO-CV) with a square error criterion, an unbiased criterion that has a closed formula for kernel ridge regression~\cite{Sellamanickan:NC2001}.

For the spherical wave function expansion, two different forms of optimization for $\lambda_\mathrm{SWF}$ were performed. The first criterion, henceforth called \textbf{SWF}, was analogous to the kernel: over the same grid, we performed LOO-CV, which also has a closed formula for quadratic problems such as~\eqref{eq:SWF_opt}~\cite{James:AnIntroductionToStatisticalLearning2023}. The second criterion took the regularization values that minimized the $\mathrm{NMSE}$ on the test points, meaning this evaluation is conditioned to the testing data. As this estimation is the best performance the model could have under our parameter choices, it was labelled \textbf{SWF (ideal)}.

Lastly, the point neuron network \textbf{PNN} was optimized using~\cite{Dixit:Zenodo2023, Zygote.jl-2018} like \textbf{Proposed}. The centers were initialized with $N_\mathrm{PNN} = 100$ centers chosen randomly with a uniform distribution in $\Omega_\mathrm{S}$, while the weights were initialized with a complex Gaussian $\mathcal{N}_\mathbb{C}(0, 1)$. The optimization constrained the centers such that $\|\mathbf{v}_n\| < R_\mathrm{in}\ \forall 1\leq n \leq N_\mathrm{PNN}$. The regularization coefficient was chosen empirically to be $\lambda_\mathrm{PNN} = 10^{-2}$.

\subsection{Experimental results}
The NMSE results, as seen in Fig.~\ref{fig:NMSE}, show that \textbf{Proposed} achieves lower estimation error overall, with $1.94~\mathrm{dB}$ lower on average than the second best, $\mathbf{PNN}$. \textbf{Proposed} also performed $2.06~\mathrm{dB}$ better than \textbf{SWF (ideal)} on average despite not having access to the test data information. Furthermore, \textbf{Proposed} and \textbf{PNN} both have similarly consistent performance over both types of arrays compared to \textbf{SWF} and \textbf{SWF (ideal)}. In addition, \textbf{Proposed} has sizably lower errors for the majority of frequencies below $1.6~\mathrm{kHz}$, where the gap between its average performance and the \textbf{PNN} widens to $2.83~\mathrm{dB}$.

\textbf{Proposed} also outperformed competing methods in trying to reconstruct the ground truth in pointwise estimates, as shown in Fig.~\ref{fig:VIS}. \textbf{SWF} shows a fading pattern as it distances from the center, while \textbf{PNN} and \textbf{Proposed} maintain stable gain throughout the region. When analyzing the NSE in Fig.~\ref{fig:NSE}, we can also see that \textbf{Proposed} not only has larger regions with lower errors, but its areas of high error are also smaller and less pronounced than \textbf{SWF} or \textbf{PNN}. This indicates that not only is \textbf{Proposed} more accurate on average throughout $\Omega_\mathrm{T}$, as indicated by the $\mathrm{NMSE}$ measurements in Fig.~\ref{fig:NMSE}, it is also a more consistent interpolation method.

\section{Conclusion}
We have proposed an exterior sound field interpolation method based on Gaussian process regression. By creating a parametric criterion based on RKHS theory, the training of our kernel is capable of adapting the estimation to the data, appropriately attenuating every order automatically.

This kernel model outperformed competing methods in simulated experiments both under spherical and random microphone distributions. This is initially evidenced in aggregate terms, with the proposed method showing lower normalized mean square error values for the majority of frequencies between $100~\mathrm{Hz}$ and $2.5~\mathrm{kHz}$. The proposed method also showed higher pointwise consistency in reconstructing the ground truth at $1~\mathrm{kHz}$, with lower and better distributed errors on each point of the analyzed region.

 \vfill
\pagebreak

\bibliographystyle{IEEEtran.bst}
\bibliography{str_def_abrv,sklab_en,refs}


\end{document}